# Influence of Mobility Restrictions on Transmission of COVID-19 in the state of Maryland - the USA

Nandini Raghuraman[1], Dr. Kartik Kaushik[1]

[1]Department of Epidemiology and Public Health University of Maryland Baltimore School of Medicine

1. Abstract:

Background: The novel coronavirus, COVID-19, was first detected in the United States in January 2020. To curb the spread of the disease in mid-March, different states issued mandatory stay-at-home (SAH) orders. These nonpharmaceutical interventions were mandated based on prior experiences, such as the 1918 influenza epidemic. Hence, we decided to study the impact of restrictions on mobility on reducing COVID-19 transmission.

Methods: We designed an ecological time series study with our exposure variable as Mobility patterns in the state of Maryland for March- December 2020 and our outcome variable as the COVID-19 hospitalizations for the same period. We built an Extreme Gradient Boosting (XGBoost) ensemble machine learning model and regressed the lagged COVID-19 hospitalizations with Mobility volume for different regions of Maryland.

Results: We found an 18% increase in COVID-19 hospitalizations when mobility was increased by a factor of five, similarly a 43% increase when mobility was further increased by a factor of ten.

Conclusion: The findings of our study demonstrated a positive linear relationship between mobility and the incidence of COVID-19 cases. These findings are partially consistent with other studies suggesting the benefits of mobility restrictions. Although more detailed approach is needed to precisely understand the benefits and limitations of mobility restrictions as part of a response to the COVID-19 pandemic.

2. Background:

An outbreak of pneumonia with an initial unknown cause was detected in Wuhan, in central China, in December 2019[1]. Later the cause for this disease was identified as the novel severe acute respiratory syndrome coronavirus 2, SARS-CoV-2[2]. After the first few reported cases, the outbreak quickly spread across the globe in two months, leading the World Health Organization to declare Coronavirus Disease 2019 (COVID-19) as a pandemic. The virus's infection was determined to range from asymptomatic to mild symptoms such as cough, shortness of breath, fever, and respiratory discomfort to severe respiratory illness and death. The SARS-CoV-2 virus's primary mode of transmission was through small droplets and aerosols from coughing or sneezing, and it spread quickly through close contact.[3].

Various mathematical models on transmission simulations were built from prior experiences, such as the 1918 influenza epidemic [4, 5]. One of the most widely implemented nonpharmaceutical interventions to control the contagion was imposing restrictions on mobility[6].

In China, a nationally coordinated effort restricting travel and social interaction along with rigorous contact tracing and testing effectively mitigated the disease's spread [7, 8]. Specifically, in a study conducted in Shenzhen, China, investigators found that restricting the mobility of the population by 20-60% helped reduce the number of new cases by 33%. Here the mobility of the population was quantified using mobile global positioning system data[9]. Similarly, in Italy, one of the first countries in Europe to experience the widespread transmission of COVID-19, a study was conducted to access the population's mobility habits and their impact on the spread of COIVD-19. The study found a prominent relationship between new incident cases and corresponding trips made three weeks before. Here the mobility of the population was quantified using car traffic count automatic sensors data[10].

Compared to nationally mandated restrictions by several countries, the United States' stay-at-home order and temporary shutdown of non-essential businesses and schools were subjected to variable levels of enforcement amongst different states resulting in a highly varying outbreak mitigation response. Due to this, varying intensities of the outbreak were observed around the country, with some counties nearing their peak while others remained in the early stages of the pandemic[11]. While the reduction in social contact and increase in physical distancing has proven to decrease infection spread[12] these results were from the initial stages of the pandemic. Since then, we have observed fluctuating mobility restrictions and disease incidence [13-15]. These variations in mobility restrictions pose a challenge in evaluating mobility restrictions' effectiveness in controlling the spread of COVID-19.

To understand how well mobility restrictions have helped control the pandemic over the past year, we will model the infection rates and aggregated mobile device location data among Maryland counties. In addition, we aim to show that mobility restrictions have impacted the spread of COVID-19 over the year and help determine different levels required to be effective given the origin of new SARS-CoV-2 strains.

3. Methods:

3.1 Study Population and Data
We performed an ecological time series study modeling the impact of mobility restrictions on COVID-19 transmission among Maryland counties. We procured the daily county-level mobility data from third-party data providers who processed and integrated mobile device locations from anonymously opted-in individuals from January till December 2020. The exposure variable data consisted of origin and destination information in zip-codes of all observed trips. We classified those trips whose origin and destination were not in the same county as mobility inflow for the destination county. The COVID-19 incidence data was obtained from the Maryland Institute for Emergency Medical Services Systems (MIEMSS) Facility Resource Emergency Database (FRED). The MIEMSS under the state's regulation coordinated the statewide emergency medical services and grouped the counties as regions based on the demand for medical service. Table 1 shows the Maryland counties grouped as regions by MEIMSS. In addition, the FRED contains the daily report of the outcome variable from all hospitals on their current COVID-19 case status for March till December 2020.

Table 1: Maryland Counties grouped into Regions by MEIMSS

| Region I | Region II | Region III | Region IV | Region V |
|---|---|---|---|---|
| Garrett County Allegany County | Frederick County Washington County | Anne Arundel County Baltimore County Baltimore City Carroll County Harford County Howard County | Worcester County Kent County Dorchester County Talbot County Somerset County Queen Anne's County Cecil County | Prince George's County Charles County St. Mary's County Calvert County Montgomery County |

3.2 Statistical Analysis

3.2.1 Cross-Correlation between the time series
We performed a cross-correlation analysis to determine any association between the volume of mobility and COVID-19 incidence amongst the counties of Maryland. The highest correlation coefficient's corresponding lag value represented the duration by which the mobility time series preceded the COVID-19 incidence.

3.2.2 Developing and Evaluating the Prediction Model
In order to build the prediction model of the trends, we employed Extreme Gradient Boosting (XGBoost) [16] an ensemble machine learning method capable of performing both regression and classification analysis. The XGBoost algorithm combines the results from multiple weak predictions to derive an accurate final prediction. The algorithm is built in a way to prevent overfitting since it combines multiple predictions automatically. We used the lag factor from our cross-correlation of the time series and created a data matrix with lagged values of the COVID-19 Hospitalizations. The model's performance was then evaluated first by splitting this data matrix into training and testing sets. The training set consisted of total hospitalizations due to COVID 19 and Total Mobility for each region from March 25, 2020, till September 23, 2020. The testing set consisted of the same but from September 24, 2020, till December 21, 2020. The algorithm was trained using the training set and made predictions on the testing set. Since this technique data matrix into training and testing sets has high variance, it impacts the algorithm's accuracy of predictions. Hence, we performed cross-validation to evaluate the performance of the algorithm further. In this approach, the dataset was split into k-parts, and each split was called the fold. The algorithm is trained on k-1 folds, where one is held to test on the held back fold. This process is continued until each fold was tested on the held back fold. Since our dataset contains more than tens of thousands of observations, a k value of 3 was employed to test the algorithm's performance. The algorithm's accuracy was quantified using common indexes, namely the rooted mean squared error (RMSE) and mean absolute percentage (MAPE), which are, respectively, defined as follows:

$$\text{RMSE} = \sqrt{\frac{1}{n}\sum_{i=1}^{n}(Predicted\ value - Observed\ value)^2} \quad (1)$$

$$\text{MAPE} = \frac{1}{n}\sum_{i=1}^{n}\left|\frac{Predicted\ value - Observed\ value}{Observed\ value}\right| \times 100\% \quad (2)$$

3.2.3 Sensitivity Analysis

We performed sensitivity analysis by increasing the total mobility volume in the testing set alone by a factor of five and then by a factor of ten. This change in mobility helped us predict the percentage increase in COVID-19 hospitalizations from the observed COVID-19 hospitalizations. The analyses were performed using R (version 4.0.0, R Foundation for Statistical Computing, Vienna, Austria).

4. Results:

The prevalence of COVID-19 cases in acute care and ICU for March – December 2020 was relatively lower for regions I and II with $8.43 \times 10^{-3}$ and $3.89 \times 10^{-3}$ prevalence, respectively, compared to regions III, IV, and V with $24.98 \times 10^{-3}$, $18.89 \times 10^{-3}$, and $17.66 \times 10^{-3}$ prevalence respectively (Figure 1). This difference in prevalence is noticed because regions III, IV, V have densely populated counties compared to regions I and II. The Association between mobility and hospitalization due to COVID-19 was assessed using cross-correlation, and the highest correlation coefficient was 0.36, and its corresponding lag factor was 18 days (Fig 3). This indicates that any change in mobility pattern takes 18 days to see any change to the COVID-19 hospitalization pattern. This aligns with the latent period for the SARS-CoV-2 virus, roughly ranging from 14-21 days. The XGBoost model's COVID-19 hospitalizations prediction accuracy was measured by the rooted mean squared error (RMSE) and mean absolute percentage error (MAPE). These accuracy indices were computed for each region and for the whole state of Maryland. For regions I and II, the model had an RMSE of 5.65 and 7.49 respectively and a MAPE of 0.10 and 0.11 respectively, while for regions III, IV, V, the Model had an RMSE of 24.96, 5.55, and 16.81 respectively and a MAPE of 0.04, 0.086 and 0.049 respectively. For the state of Maryland, the Model's RMSE was 38.48, and the MAPE was 0.036 (Fig 4). Lower the RMSE and MAPE scores indicate higher prediction accuracy. As we can see, regions I, II, and IV have low MAPE and RMSE scores.

In contrast, regions III, V, and the whole state of Maryland had slightly higher RMSE values and similar MAPE values, indicating that the forecast model had relatively high accuracy in predicting the COVID-19 hospitalizations for regions that had a relatively lower number of COVID-19 cases. From our sensitivity analysis, we can see increased mobility leading to more hospitalizations (Fig5). As the mobility was increased by a factor of five, the percentage increase in COVID-19 hospitalizations was 18%; similarly, when the mobility was increased by a factor of ten, the percentage increase in COVID-19 hospitalizations was 43%. From a health guideline perspective, this was one of the main reasons people were advised to travel only for essential purposes.

5. Discussion:

The findings of our study demonstrated a positive linear relationship between mobility and incidence of COVID-19 cases as we saw an 18-43% increase in the incidence of COVID19 cases as mobility was increased by a factor of five and ten. These findings are partially consistent with other studies suggesting the benefits of mobility restrictions[14]; for example, an earlier study of government-mandated physical distancing intervention demonstrated an association with a more significant reduction in COVID-19 incidence. In addition, another study conducted in China modeled the COVID-19 incidence using mobile phone data and provided further support for this view. However, the study was conducted during the early stage of the pandemic [9, 17].

Our study uses the daily mobility data from a third-party data provider who processed and integrated mobile device location from anonymously opted-in individuals. Using a similar mobile device tracking method, a study in the United States found a strong correlation between decreased mobility and lower COVID-

19 case incidence [18]. In addition, although our study did not factor in the inter-household spread, there have been other studies investigating mobility in residential areas but found it challenging to interpret this phenomenon of inter-household mixing[19]. Finally, our finding of lag time as 18 days is consistent with prior research suggesting a typical lag between these changes in mobility patterns and the onset of the disease showing symptoms to be around 14-21 days [18, 20, 21].

Mandatory restrictions on mobility may reduce the frequency of interpersonal interactions. As these restrictions are lifted, precautionary behaviors such as social distancing may have become normalized, reducing the effects of increased mobility. When this is combined with better quarantine measures and contact tracing, the relative benefits of mobility restrictions might decrease as the population adapts to other risk mitigation measures [17]. While our finding supports the role of mobility restrictions as a critical strategy to curb the pandemic, it is still unclear to suggest a scale-up of other coordinated mitigation strategies simultaneously to drive transmission to a lower extent and maximize health gains.[22-26]

This study has several limitations. First, although the mobility data from a third-party data provider weighted for those individuals who did not have their mobile device location anonymously opted-in, there is a chance that we could not have accounted for the floater population of the state of Maryland. Second, the use of an ecological study design is prone to confounding. Our data do not allow us to isolate the impact of mobility restrictions from the other nonpharmacological interventions mitigated as a pandemic response, such as the use of face coverings, adherence to social distancing guidelines, degree of inter-household mixing, and the availability of rapid testing and controlling an outbreak. However, from an analytics perspective disentangling these factors would be challenging given the complex nature of these relationships even if data were available. Third, limitation concerning generalizability as in the United States, there were sub-national differences in implementing these restriction guidelines. The final limitation of this study is the use of a single machine learning model to make predictions as the accuracy of prediction of the current model was not compared to other traditionally used machine learning models for time series forecasting, such as the autoregressive integrated moving average model (ARIMA).

Our analysis somewhat extends the understanding of the dynamics when mobility is restricted in response to a pandemic. An accurate forecast model with real-time mobility data will forecast the number of hospitalizations and help estimate the demand vs. capacity for hospitals and clinics. This would help us stay ahead of any increases in hospitalizations. Further incorporating other factors, we would be able to estimate the peak demand for hospital beds, the time scale to ramp up facilities, the timeline for the peak to last, and the subsequent decrease. Also, evaluating various scenarios and getting an approximate idea of potential requirements would influence policy decisions. Further combining with the rate of recovery, average days to recover, and other similar data, we will be able to develop a comprehensive plan of action to mitigate the risks of the pandemic.

6. Conclusion

Restrictions on mobility, or lockdowns, are necessary instruments to curb the pandemic at an initial stage. Last year, these stringent restrictions bought time to establish effective testing and tracking systems, but they did have a considerable economic cost. Synergistically, restrictions on mobility and other nonpharmacological interventions such as wearing face coverings and social distancing have made it possible to slowly open society for regular functioning. While additional work is needed to enhance these findings, they may be helpful to policymakers as they determine the importance of tracking the trends in mobility when developing comprehensive strategies.

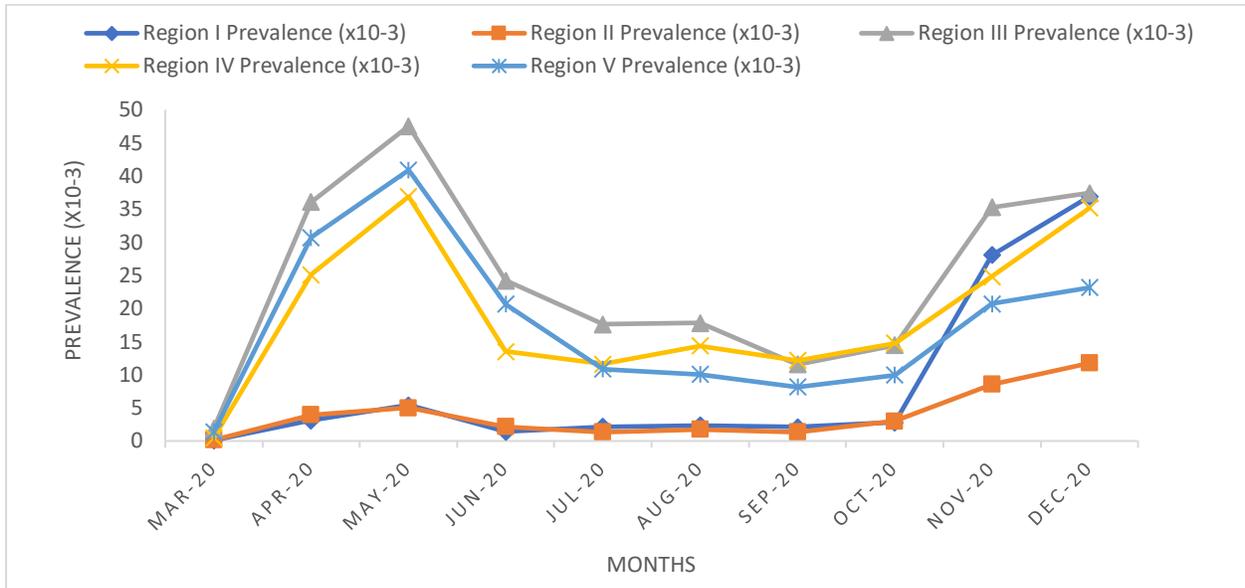

Figure 1: Monthly breakdown of COVID-19 Patients in Acute Care and ICU by Region

Fig 1: The prevalence of COVID-19 patients in Acute care and ICU was determined for each month starting from March 2020 till December 2020 by aggregating the total number COVID-19 cases for each month and dividing by the mean population of that region.

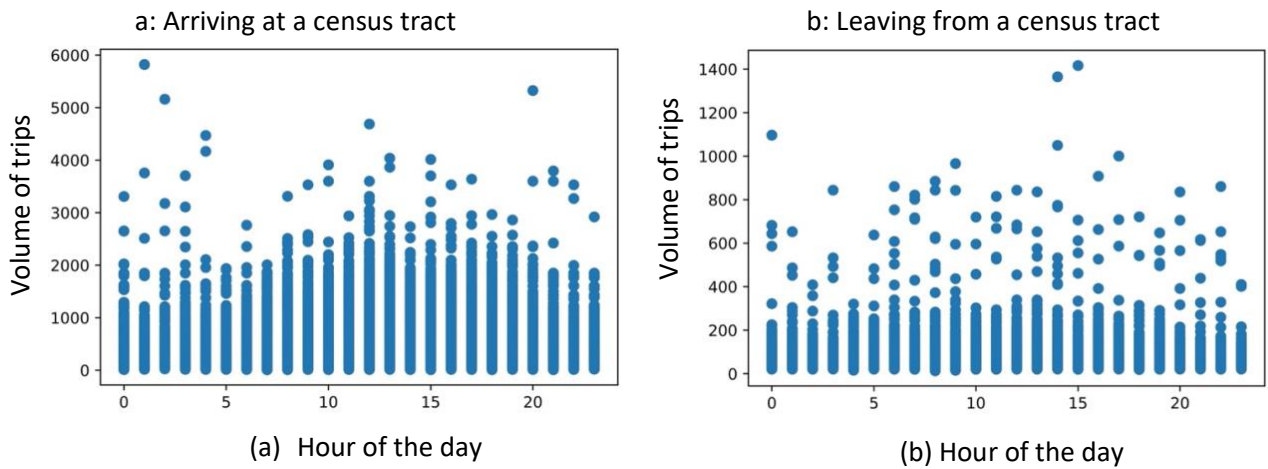

Figure 2a&b: Volume of trips arriving and leaving from a randomly chosen census tract

a: Arriving at a census tract

b: Leaving from a census tract

(a) Hour of the day

(b) Hour of the day

Figure 3: Cross-Correlation between Mobility and COVID-19 Hospitalizations

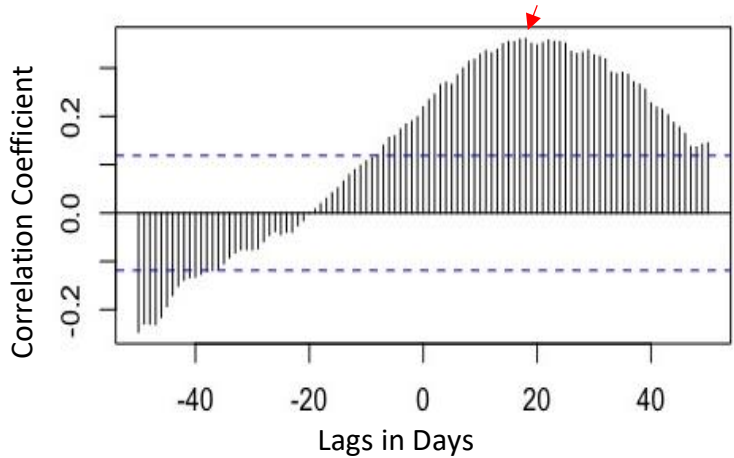

Fig 3: The cross correlation between the Mobility time series and COVID-19 hospitalizations time series had highest correlation coefficient of 0.36 with a corresponding lag of 18 days.

Figure 4: XGBoost Model prediction of COVID-19 Hospitalizations

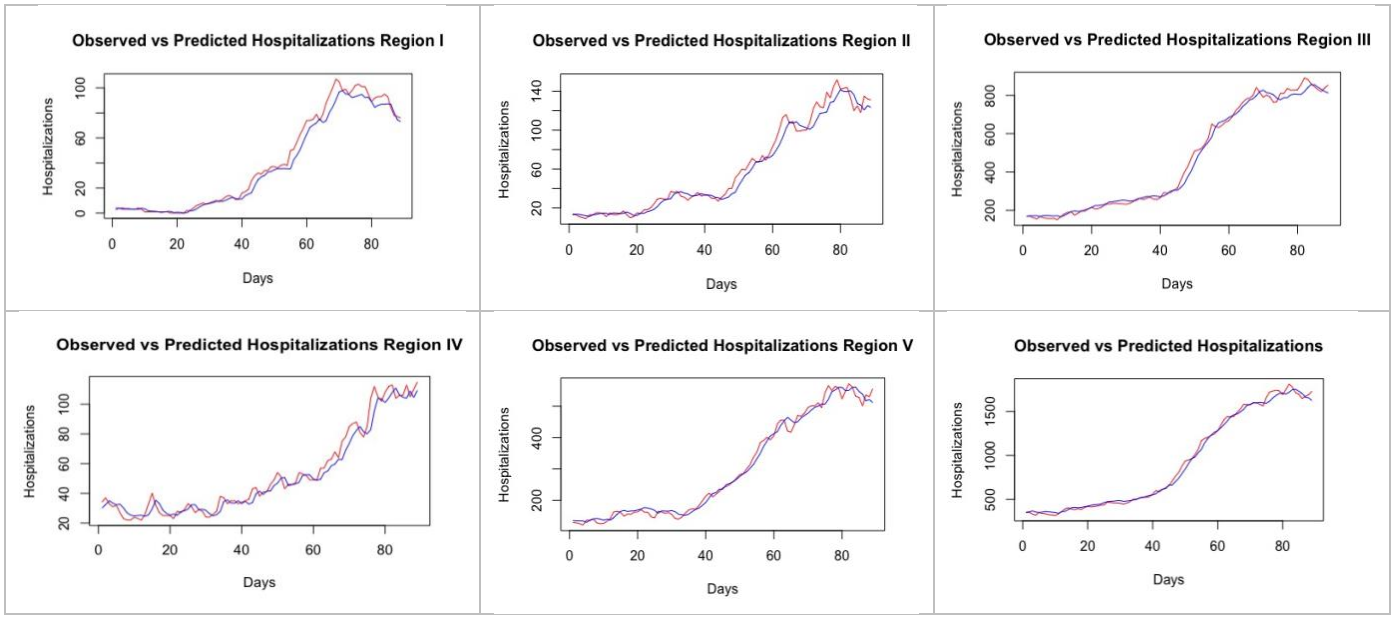

Fig 4: The line graphs represent the actual vs predicted COVID-19 hospitalizations, where the actual COVID-19 hospitalizations is represented by the red-line and the predicted COVID-19 hospitalizations is represented by the blue line.

Figure 5: XGBoost model predicted COVID-19 Hospitalizations by factor increase in mobility

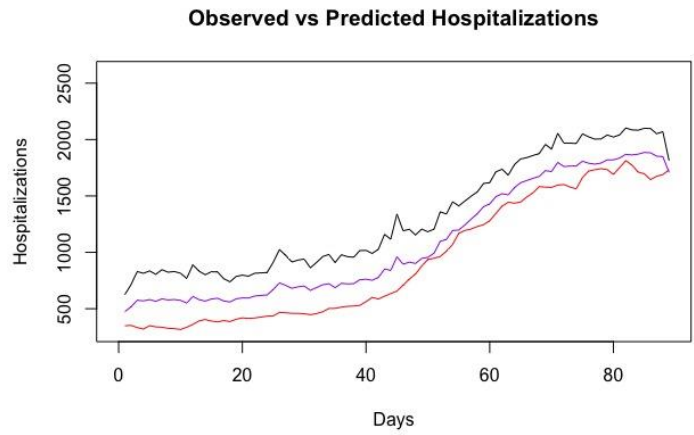

Fig 5: The line graph represents the actual vs predicted hospitalizations due to COVID-19, where the actual COVID-19 hospitalizations is represented by red-line and the predicted COVID-19 hospitalizations is represented by purple line when the mobility is increased by a factor of five. The black line represents the predicted COVID-19 hospitalizations when the mobility is increased by a factor of 10.